\documentclass{article}
\usepackage[utf8]{inputenc}

\usepackage{arxiv}

\usepackage[utf8]{inputenc} % allow utf-8 input
\usepackage[T1]{fontenc}    % use 8-bit T1 fonts
\usepackage{hyperref}       % hyperlinks
\usepackage{url}            % simple URL typesetting
\usepackage{booktabs}       % professional-quality tables
\usepackage{amsfonts}       % blackboard math symbols
\usepackage{nicefrac}       % compact symbols for 1/2, etc.
\usepackage{microtype}      % microtypography
\usepackage{lipsum}		% Can be removed after putting your text content
\usepackage{graphicx}
\usepackage[numbers]{natbib}
\usepackage{doi}
\usepackage{float} 
\usepackage{comment}
\usepackage[linesnumbered,ruled,vlined]{algorithm2e}

\usepackage{acro}

\DeclareAcronym{rvqc}{
  short=RVQC,
  long=Recursive Variational Quantum Compiling,
}

\DeclareAcronym{vqc}{
  short=VQC,
  long=Variational Quantum Compiling,
}
\DeclareAcronym{fumc}{
  short=FUMC,
  long=Full Unitary Matrix Compiling,
}

\DeclareAcronym{fisc}{
  short=FISC,
  long=Fixed Input State Compiling,
}

\DeclareAcronym{let}{
  short=LET,
  long=Loschmidt Echo Test,
}

\DeclareAcronym{opr}{
  short=OPR,
  long=Optimal Parameter Resilience,
}

\DeclareAcronym{nisq}{
  short=NISQ,
  long=Noisy Intermediate-Scale Quantum,
}

\SetCommentSty{mycommfont}

\SetKwInput{KwInput}{Input}                % Set the Input
\SetKwInput{KwOutput}{Output}  

\usepackage{amsmath,amssymb}  %I added this so that you can use the align tool for equations!
\usepackage{physics}
\usepackage{bm}

\input{qcircuit.sty}

\usepackage[colorinlistoftodos]{todonotes}

\title{Recursive Variational Quantum Compiling}

\author{
  Stian Bilek \\
  Department of Physics \\
  University of Oslo \\
  Oslo, Norway\\
  \texttt{stianbilek@gmail.com} 
  \\
  \texttt{stian.bilek@fys.uio.no} \\
   \And
  Kristian Wold \\
  Department of Computer Science  \\
  Oslo Metropolitan University \\
  Oslo, Norway\\
  \texttt{krisw@oslomet.no} \\
}

\hypersetup{
pdftitle={Recursive Variational Quantum Compiling},
pdfauthor={Stian Bilek, Kristian Wold},
}

\begin{document}
\maketitle

\begin{abstract}
    Variational quantum compiling (VQC) algorithms aim to approximate deep quantum circuits with shallow parameterized ansatzes, making them more suitable for NISQ hardware. In this article a variant of VQC named the recursive variational quantum compiling (RVQC) algorithm is proposed. Existing VQC algorithms typically require coherently executing the full circuit during compilation. Under the influence of noise, sufficiently deep target circuits make compiling unfeasible using ordinary VQC. Since the compiling is often accomplished using a gradient-based quantum-classical approach, the quantum noise manifest as a noisy gradient during optimization, making convergence hard to obtain. On the other hand, RVQC can compile a circuit by first dividing it into $N$ shorter sub-circuits, then evaluate one sub-circuit at a time. As a result, the circuit depth required to implement RVQC is not dependent on the depth of the target circuit, but on the depth of the sub-circuits. Choosing a high enough $N$ thus ensures sufficiently shallow sub-circuit which can be successfully compiled individually. We show mathematical evidence of this property. RVQC was compared with VQC on a noise model of the IBM Santiago device with the goal of compiling several randomly generated five-qubit circuits of approximately depth 1000. It was shown that VQC was not able to converge within 500 iterations of optimization. On the other hand, RVQC was able to converge to a fidelity of $0.90 \pm 0.05$ within a total of 500 iterations when splitting the target circuits into $N = 5$ parts. We argue that this comes as a result of the mitigation of noise-induced barren plateaus.
\end{abstract}

\section{Introduction}
Quantum computers show potential in solving problems outside the reach of classical computers \cite{365700,doi:10.1126/science.273.5278.1073}. Unfortunately, today's noisy intermediate-scale quantum (NISQ) devices have quite limited coherence times. We are hence dependent on algorithms that are not only efficient, but also have a shallow circuit depth (the number of unique time steps during which gates are applied \cite{NielsenChuang2010}), in order to respect the specific hardware \cite{Preskill2018quantumcomputingin}. Hybrid quantum-classical algorithms are perfect candidates for NISQ devices, as they are built on the philosophy that even minimal quantum resources could be made useful when used in conjunction with classical routines \cite{McClean_2016}. Examples of such algorithms are the variational quantum eigensolvers (VQE) for finding the ground state of a Hamiltonian \cite{VQEog}, the quantum approximate optimization algorithm used for optimization problems \cite{farhi2014quantum} and parameterized quantum circuits for machine learning problems \cite{Benedetti_2019}. Even though these algorithms are promising, could one utilize the hybrid quantum-classical approach to run existing deep quantum circuits on NISQ devices? 

A general approach for making deep circuits more suitable for NISQ devices is to reduce their depth using \emph{Variational Quantum Compiling} (VQC) \cite{qaqc}. This approach aims to approximate a deep circuit $U$ using a shallow parameterized circuit $V(\boldsymbol{\theta})$, where $\boldsymbol{\theta}$ are the trainable parameters. The parameters are trained by minimizing a cost function in a hybrid quantum-classical fashion, where the cost usually is defined using some measure of distance between $U$ and $V(\boldsymbol{\theta})$. \citet{noiseresiliencevqc} showed that VQC exhibit  \emph{Optimal Parameter Resilience} (OPR), meaning that under typical incoherent noise models, such as decoherence processes and readout errors, the cost function has the same minimum as the cost function in the ideal scenario. Moreover, their numerical experiments indicate that this property also holds for more realistic noise models. This suggests that it could possible to learn optimal, or at least approximately optimal, parameters when compiling circuits with VQC on noisy quantum hardware. 

While the cost functions may have the same minimum in the ideal and noisy scenario, \citet{noiseresiliencevqc} does not address the difficulty of finding this minimum, especially as the circuit depth and amount of noise increase. In this article, we illustrate that too deep circuits can be difficult to compile when noise is introduced. Even though they are easily compiled in the ideal scenario, the introduction of large amounts of noise causes gradient descent to essentially perform random steps in parameter space, resulting in a failure to converge. This effect is closely related to noise-induced barren plateaus, as demonstrated in \citet{wang2021noise}. Their work rigorously proves that local Pauli noise in variational quantum algorithms (VQAs) leads to an exponential decay of gradients with increasing qubit count. Unlike barren plateaus arising from random initialization in noise-free settings, these noise-induced barren plateaus are a result of the exponential suppression of gradients due to local Pauli noise, which is yet another limiting factor for the scalability of variational quantum algorithms. To amend this, we introduce a recursive variational quantum compiling algorithm (RVQC) that is able to successfully approximate such deep circuits despite of the noise. The algorithm works by dividing the target circuit $U$ into $N$ parts and recursively compiles one part at a time until the whole of $U$ is approximated by a single ansatz $V(\boldsymbol{\theta})$. The circuit depth required to run the algorithm is not dependent on the depth of the target circuit, but on the depth of each of the $N$ parts and on the depth of the parameterized ansatzes.
For a shallow ansatz and large $N$, the effective noise in each resulting circuit is greatly reduced compared to the full circuit. This allows gradient descent to perform informed updating steps and converge to an approximation greatly outperforming the standard VQC approach. To illustrate the algorithm, we will attempt to compile several five-qubit random circuit using both ordinary VQC and RVQC on simulated noisy hardware. We then calculate the density matrices of the resulting optimized ansatzes and compare them with the density matrix of the full circuits.

In section \ref{sec:theory}, we will explain VQC and RVQC in detail. We will then go through the target circuits, variational ansatzes and optimization methods used to test out the algorithm in section \ref{sec:method}. We present and discuss the results in section \ref{sec:results}, while we finally conclude the study and propose further work in section \ref{sec:Conclusion}. Supplementary material is provided in the appendix.

\printacronyms[display=all]

\section{Theory}
\label{sec:theory}
\subsection{Variational Quantum Compiling}
Variational quantum compiling (VQC) is a method for compiling quantum circuits, i.e., finding approximate circuits smaller than the original, by optimizing a variational ansatz $V(\boldsymbol{\theta})$. This can be achieved by minimizing the following cost function with respect to the circuit parameters $\boldsymbol{\theta}$ \cite{qaqc}:

\begin{align}
    \label{eq:FUMC}
    C_{FUMC}(\boldsymbol{\theta}) = 1 - |Tr(V^{\dagger}(\boldsymbol{\theta})U)|^2/d^2,
\end{align}
where the last term can be identified as the Hilbert-Schmidt inner product of $V(\boldsymbol{\theta})$ and $U$, and $d=2^n$ is the dimension of the Hilbert space for $n$ qubits. This particular loss yields a Full Unitary Matrix Compiling (FUMC), since the expression is minimized if and only if  $V(\boldsymbol{\theta}) = U$ up to a global phase. This cost function typically require many gates and a number of ancilliary qubits, making it generally expensive and unsuitable for NISQ hardware \cite{qaqc}. Alternatively, one can perform a Fixed Input State Compiling (FISC) where one is only interested in the effect of $U$ on a particular input state $\ket{\psi_0}$ \cite{qaqc}. This input state can be taken to be the all-zero state, i.e., $\ket{\psi_0} = \ket{\boldsymbol{0}}$, the typical initial state prepared by quantum computers. We can implement FISC by minimizing the following cost function with respect to the parameters $\boldsymbol{\theta}$:

\begin{align}
    \label{eq:FISC}
    C_{FISC}(\boldsymbol{\theta}) = 1 - |\bra{\boldsymbol{0}} U^{\dagger} V(\boldsymbol{\theta})\ket{\boldsymbol{0}}|^2,
\end{align}
where the last term can be identified as the Loschmidt Echo Test (LET) \cite{Goussev:2012}. For $C(\boldsymbol{\theta}) \approx 0$, we have that $V(\boldsymbol{\theta})\ket{\boldsymbol{0}} \approx U\ket{\boldsymbol{0}}$. This is a more restricted form of compiling compared to FUMC, but could in turn be more suitable for noisy quantum hardware since the trained ansatzes generally consist of fewer gates and no ancilla qubits. In this article, the focus will be on FISC because it is more suitable for noisy hardware. However, note that the following discussion can easily be extended to FUMC by introducing the appropriate cost function.

\subsection{Recursive Variational Quantum Compiling}
Since quantum circuits are sequences of quantum gates, they can be written as a product of sub-sequences of gates, i.e., sub-circuits. Therefore, we can decompose a given quantum circuit $U$ as

\begin{align}
\begin{split}
    \label{eq:unitarymatrixproduct}
    U = U_N U_{N-1} \cdots U_2 U_1,\\
    U^{(k)} = U_k U_{k-1} \cdots U_2 U_1
\end{split}
\end{align}
where $N$ is the number of sub-circuits $U_k$, and $U^{(k)}$ consists of the first $k$ sub-circuits. See figure \ref{circuit:UdividedCircuit} for an example with $N = 3$.

%\begin{equation}
%    \hat{H}_T(\boldsymbol{\omega}) = %\sum_i \omega_i \hat{h}_i
%\end{equation}

%\begin{equation}
%  \ket{\psi(t)} =   %e^{-i\hat{H}_T(\boldsymbol{\omega})t}\ke%t{\psi_0}
%\end{equation}

%\begin{equation}
%    E = \bra{\psi(t)} \hat{H}_E %\ket{\psi(t)}
%\end{equation}

%\begin{equation}
%    \frac{\partial E}{\partial \omega_k}
%\end{equation}

\begin{figure}[h]
\[U \equiv \begin{array}{c}
\Qcircuit @C=1em @R=1em {
& \gate{H} & \ctrl{1} & \gate{X} & \qswap & \gate{R_y(\pi/3)} & \targ & \qw  \\
& \gate{H} & \targ & \gate{Y} & \qswap \qwx & \qw  & \ctrl{-1} & \qw \gategroup{1}{2}{2}{3}{.7em}{--} \gategroup{1}{4}{2}{5}{.7em}{--} \gategroup{1}{6}{2}{7}{.7em}{--} \\
}
\end{array}\]
$$\qquad \underbrace{\qquad \qquad}_{U_1}     \underbrace{\quad \qquad }_{U_2} \underbrace{\qquad \qquad \qquad \quad }_{U_3}$$
\caption{Quantum circuit divided into $N = 3$ different subcircuits, each representable with some unitary operator.}
\label{circuit:UdividedCircuit}
\end{figure}
The Recursive Variational Quantum Compiling (RVQC) algorithm performs FISC, with $\ket{\boldsymbol{0}}$ as input state, on quantum circuits represented by  \autoref{eq:unitarymatrixproduct}. Rather than minimizing a single cost function to approximate the whole of $U$ at once, RVQC learns to approximate the circuit one sub-circuit at a time. The first sub-circuit $U_1$ of the circuit can be approximated by an ansatz $V(\boldsymbol{\theta}^{(1)})$ by minimizing the usual LET cost function

\begin{align}
    \label{eq:CostFunction1}
    C(\boldsymbol{\theta}^{(1)}) = 1 - |\bra{\boldsymbol{0}} U_1^{\dagger} V(\boldsymbol{\theta}^{(1)})\ket{\boldsymbol{0}}|^2,
\end{align}
where $\boldsymbol{\theta}^{(1)}$ are the trainable parameters of the ansatz. For $C(\boldsymbol{\theta}^{(1)}) \approx 0$, we have that $V(\boldsymbol{\theta}^{(1)})\ket{\bold{0}} \approx U_1\ket{\boldsymbol{0}}$. The next subcircuit, $U_2$, can be incorporated by minimizing the cost function

\begin{align}
    \label{eq:CostFunction2}
    C(\boldsymbol{\theta}^{(2)}) = 1-|\bra{\boldsymbol{0}} V^{\dagger}(\boldsymbol{\theta}^{(1)})U_2^{\dagger}V(\boldsymbol{\theta}^{(2)}) \ket{\boldsymbol{0}}|^2,
\end{align}
where $\boldsymbol{\theta}^{(2)}$ is a new set of parameters independent of $\boldsymbol{\theta}^{(1)}$. Hence, the ansatz $V(\boldsymbol{\theta}^{(2)})$ is able to approximate the product of $U_1$ and $U_2$ without having to coherently execute both. This gives rise to the RVQC algorithm, which involves iteratively minimizing the following cost functions for $k \in \{1, \cdots, N\}$:

\begin{align}
    \label{eq:CostFunction}
    C(\boldsymbol{\theta}^{(k)}) = 1-|\bra{\boldsymbol{0}} V^{\dagger}(\boldsymbol{\theta}^{(k-1)})U_k^{\dagger}V(\boldsymbol{\theta}^{(k)}) \ket{\boldsymbol{0}}|^2,
\end{align}
where $V(\boldsymbol{\theta}^{(0)}) = I$. After minimizing all the cost functions, the final ansatz $V(\boldsymbol{\theta}^{(N)})$ will approximatly encode the effect of the whole circuit $U$ acting on $\ket{\boldsymbol{0}}$. This scheme bypasses the need to evaluate all of $U = U_N U_{N-1}\cdots U_1$ coherently during optimization, which can be potentially unfeasible if $U$ is a deep circuit and the hardware used is sufficiently noisy. We will show evidence that RVQC can compile deep circuits under the influence of simulated noise, where ordinary VQC would otherwise fail.

\begin{figure}[h]
\[\begin{array}{c}
\Qcircuit @C=1em @R=1em {
& \multigate{3}{V(\boldsymbol{\theta}^{(k)})} & \qw & \qw & \multigate{3}{{U^\dagger_k}} & \multigate{3}{V^\dagger(\boldsymbol{\theta}^{(k-1)})} & \meter\\
& \ghost{V(\boldsymbol{\theta}^{(k)})} & \qw & \qw & \ghost{{U^\dagger_k}} & \ghost{V^\dagger(\boldsymbol{\theta}^{(k-1)})} & \meter\\
& \nghost{V(\boldsymbol{\theta}^{(k)})} & \vdots && \nghost{{U^\dagger_k}} & \nghost{V^\dagger(\boldsymbol{\theta}^{(k-1)})} &  \meter \\
& \ghost{V(\boldsymbol{\theta}^{(k)})} & \qw &\qw & \ghost{{U^\dagger_k}} & \ghost{V^\dagger(\boldsymbol{\theta}^{(k-1)})}  & \meter \\}
\end{array}\]

\caption{Quantum circuit used to implement the cost function \autoref{eq:CostFunction}. The unitary $V(\boldsymbol{\theta}^{(k)})$ being equal (up to a global phase) to $V(\boldsymbol{\theta}^{(k-1)})U_k$ is consistent with measuring the all-zero state $\ket{\boldsymbol{0}}$, in the ideal case. We can thus minimize the cost function in \autoref{eq:CostFunction} by maximizing the probability of measuring $\ket{\boldsymbol{0}}$ in the above circuit.}
\label{fig:CostFunction}
\end{figure}

The LET cost function \autoref{eq:CostFunction} can be evaluated on a quantum computer by implementing the circuit given by \autoref{fig:CostFunction} and calculating the probability for measuring the qubits in the all-zero state $\ket{\boldsymbol{0}}$. Mathematically, this can be expressed as

\begin{equation}
\label{eq:LSET}
    C(\boldsymbol{\theta}^{(k)}) = 1 - P(\ket{\boldsymbol{0}};\boldsymbol{\theta}^{(k)}) \approx 1 - \frac{1}{n_s}\sum_{i=1}^{n_s} p_i(\boldsymbol{\theta}^{(k)}),
\end{equation}
where $P(\ket{\boldsymbol{0}};\boldsymbol{\theta}^{(k)})$ is the probability of measuring $\ket{\boldsymbol{0}}$ after applying  circuit \autoref{fig:CostFunction}, $n_s$ is the number of shots used, and $p_i(\boldsymbol{\theta}^{(k)})$ indicates whether experiment $i$ was measured to be $\ket{\boldsymbol{0}}$ or not ($p_i(\boldsymbol{\theta}^{(k)}) = 1$ or $0$, respectively).

\subsection{Preventing Exponential Gradient Decay by Splitting into Sub-Circuits}
\label{subsec:preventingExponentialGradientDecay}

\subsubsection{Noise-Induced Barren Plateaus}

Consider an $n$-qubit parameterized ansatz $V(\boldsymbol{\theta})$ of depth $K$.
The cost function for VQC is defined by
\begin{equation}
    C(\boldsymbol{\theta}) \;=\; 1 \;-\; \bigl|\langle 0 \mid U_{\mathrm{target}}^\dagger\,V(\boldsymbol{\theta}) \mid 0\rangle \bigr|^2,
\end{equation}
where $U_{\mathrm{target}}$ is the target circuit, of depth $L$. We adopt a noise model similar to \cite{wang2021noise} in which local Pauli noise channels $\mathcal{N}_j$ act on each qubit $j$. In particular, for a single-qubit Pauli operator $\sigma \in \{X, Y, Z\}$, the channel $\mathcal{N}_j$ acts as
\begin{equation}
    \mathcal{N}_j(\sigma) \;=\; q_{\sigma}\,\sigma,
    \label{eq:Nj}
\end{equation}
where $-1 < q_X, q_Y, q_Z \le 1$. The noise strength is characterized by:
\begin{equation}
    q \;=\; \sqrt{\max\{\,|q_X|,\,|q_Y|,\,|q_Z|\,\}}.
\end{equation}

Let $U_i$ (or $V_i(\boldsymbol{\theta})$) denote the noiseless channel that implements the corresponding unitary on an $n$-qubit system, and define 
\[
    \mathcal{N} \;=\; \mathcal{N}_1 \,\otimes\, \cdots \,\otimes\, \mathcal{N}_n
\]
as the $n$-qubit noise channel acting independently on each qubit. Consequently, when this noise is applied in each layer, the effectively implemented circuits become
\begin{equation}
    \widetilde{V}(\boldsymbol{\theta})
    \;=\;
    \bigl(\mathcal{N} \circ V_K(\boldsymbol{\theta})\bigr)
    \;\cdots\;
    \bigl(\mathcal{N}\circ V_1(\boldsymbol{\theta})\bigr),
\end{equation}
\begin{equation}
    \widetilde{U}_{\mathrm{target}}
    \;=\;
    \bigl(\mathcal{N} \circ U_L\bigr)
    \;\cdots\;
    \bigl(\mathcal{N}\circ U_1\bigr),
\end{equation}
yielding a noisy cost function 
$\widetilde{C}\bigl(\boldsymbol{\theta}\bigr)$.

According to the noise-induced barren plateau (NIBP) theorem~\cite{wang2021noise}, if $q < 1$ is the local noise factor per layer, then the partial derivatives of $\widetilde{C}$ which scales with the depth $L+K$ satisfy
\begin{equation}
\label{eq:NoiseInduced}
    \bigl|\partial_j \widetilde{C}_{L+K}(\boldsymbol{\theta})\bigr|
    \;\le\;
    2^{(L+K)\log_2{q}}.
\end{equation}
Since $q < 1$, this upper bound decays exponentially in $L+K$. Moreover, because $L$ typically scale at least polynomially in the system size $n$, attempting to compile $U_{\mathrm{target}}$ in one shot will lead to vanishing gradients as $n$ grows.

\subsubsection{Splitting into $N$ Sub-Circuits}

To circumvent exponential decay, we split \(U_{\mathrm{target}}\) of depth \(L\) into \(N\) sub-circuits:
\begin{equation}
    U_{\mathrm{target}} \;=\; U_N \,\cdots\, U_2\,U_1,
\end{equation}
each of depth approximately \(L/N\). In addition, from the RVQC LET cost function (Eq. \ref{eq:CostFunction}), we get an overhead of \(2K\) layers per sub-circuit. Thus a recursive compilation approach only ever needs to implement a circuit of depth at most
\begin{equation}
    2K \;+\; \frac{L}{N}
\end{equation}
at a time. Hence, from the same noise-induced barren plateau (NIBP) argument, each partial derivative is of order
\begin{equation}
\label{eq:qfactorRVQC}
    2^{\left(2K + \frac{L}{N}\right)\log_2(q)},
\end{equation}
instead of $2^{(L+K)\log_2(q)}$. Thus, by choosing $N$ such that \(2K + L/N < L + K\), we should be able to achieve a larger gradient for the RVQC approach than for VQC.

To prevent the noise from inducing a barren plateau, we require that the factor in Eq. \eqref{eq:qfactorRVQC} remain at least \(1/\mathrm{poly}(n)\). That is, we aim for a polynomial decay in the number of qubits. Equivalently, we demand
\begin{equation}
\label{eq:inequality-2K-corrected}
    2^{\left(2K + \frac{L}{N}\right)\log_2(q)}
\;\ge\;
\frac{1}{\mathrm{poly}(n)}
\quad
\Longleftrightarrow
\quad
    2K \;+\; \frac{L}{N}
\;\;\le\;
\frac{\log\bigl(\mathrm{poly}(n)\bigr)}
{\log\!\bigl(\tfrac{1}{q}\bigr)}.
\end{equation}
Since \(\mathrm{poly}(n)\) can be chosen as \(n^c\) for some constant \(c>0\), the right-hand side is on the order of \(\tfrac{\log(n)}{\log(1/q)}\). Thus one obtains
\begin{equation}
\label{eq:upperboundwithc}
    2K + \frac{L}{N}
\;\le\;
\frac{c\,\log(n)}{\log(1/q)}.
\end{equation}
Hence, as long as \(N\) is chosen sufficiently large (and \(K\) is chosen sufficiently small) the factor
\[
    2^{\left(2K + \frac{L}{N}\right)\log_2(q)}
\]
remains at least \(1/\mathrm{poly}(n)\), ensuring that the gradient does not vanish exponentially in \(n\).

Equation \ref{eq:upperboundwithc} shows that as the noise level increases (i.e., as \( q \) decreases from 1), the circuit depth of each subcircuit,
\[
2K + \frac{L}{N},
\]
must be reduced accordingly to avoid an exponentially decaying gradient. In the RVQC framework, we have the flexibility to adjust the circuit depth by varying \( N \), which is not possible in the standard VQC approach.

As an example, consider a quantum circuit with \( n = 64 \) qubits and a total circuit depth
\[
L = 64^2 = 4096.
\]
In the VQC scenario, the gradient decays exponentially with the number of qubits. In contrast, for the RVQC scenario, where we might desire the gradient to decay quadratically in \( n \), we require an upper bound on \( 2K + \frac{L}{N} \). Given a noise parameter \( q = 0.99 \), this upper bound is
\[
2K + \frac{L}{N} \leq \frac{2\log(64)}{\log(1/0.99)} \approx 827.
\]
For instance, one may partition the circuit into segments of depth
\[
\frac{L}{N} = 400,
\]
and choose an ansatz depth of
\[
K = 213,
\]
which still respects the upper bound, since $2K + L/N = 826 < 827$.

\section{Method}
\label{sec:method}
In this section, we describe the methods used in the study. We explain the structure of the variational ansatz, define the RVQC algorithm, the noise model used for simulations, and the performance metrics for evaluating the results.

\subsection{Variational Ansatz}

The ansatz used to compile the five qubit random circuit is defined as seen in \autoref{fig:fivequbitansatz}, where the $R_yR_z$ blocks are independently parameterized Pauli rotations defined as in \autoref{fig:RyRz}, and the blocks $ent$ creates entanglement using the typical ladder of CNOT-gates as showed in \autoref{fig:entanglement}.

\begin{figure}[h]
\[\begin{array}{c}
\Qcircuit @C=1em @R=1em {
& \multigate{4}{V(\boldsymbol{\theta})} & \qw\\
&        \ghost{V(\boldsymbol{\theta})} & \qw\\
&        \ghost{V(\boldsymbol{\theta})} & \qw\\
&        \ghost{V(\boldsymbol{\theta})} & \qw\\
&        \ghost{V(\boldsymbol{\theta})} & \qw\\}
\end{array}
=
\begin{array}{c}
\Qcircuit @C=1em @R=1em {
& \multigate{4}{R_y R_z} & \multigate{4}{ent} & \multigate{4}{R_y R_z} & \multigate{4}{ent} & \multigate{4}{R_y R_z} & \multigate{4}{ent} & \multigate{4}{R_y R_z} & \multigate{4}{ent} & \multigate{4}{R_y R_z}  & \qw\\
&        \ghost{R_y R_z} &        \ghost{ent} &        \ghost{R_y R_z} &        \ghost{ent} &        \ghost{R_y R_z} &        \ghost{ent} &        \ghost{R_y R_z} &        \ghost{ent} &        \ghost{R_y R_z}  & \qw\\
&        \ghost{R_y R_z} &        \ghost{ent} &        \ghost{R_y R_z} &        \ghost{ent} &        \ghost{R_y R_z} &        \ghost{ent} &        \ghost{R_y R_z} &        \ghost{ent} &        \ghost{R_y R_z}  & \qw\\
&        \ghost{R_y R_z} &        \ghost{ent} &        \ghost{R_y R_z} &        \ghost{ent} &        \ghost{R_y R_z} &        \ghost{ent} &        \ghost{R_y R_z} &        \ghost{ent} &        \ghost{R_y R_z}  & \qw\\
&        \ghost{R_y R_z} &        \ghost{ent} &        \ghost{R_y R_z} &        \ghost{ent} &        \ghost{R_y R_z} &        \ghost{ent} &        \ghost{R_y R_z} &        \ghost{ent} &        \ghost{R_y R_z}  & \qw \\}
\end{array}\]

\caption{Ansatz used for approximating the five-qubit random circuit. The $R_yR_z$ blocks are independently parameterized Pauli rotations defined as in \autoref{fig:RyRz}, and the blocks $ent$ creates entanglement using CNOT-gates as showed in \autoref{fig:entanglement}.}
\label{fig:fivequbitansatz}
\end{figure}

\begin{figure}[H]
\[\begin{array}{c}
\Qcircuit @C=1em @R=1em {
& \multigate{4}{R_y R_z} & \qw\\
&        \ghost{R_y R_z} & \qw\\
&        \ghost{R_y R_z} & \qw\\
&        \ghost{R_y R_z} & \qw\\
&        \ghost{R_y R_z} & \qw\\}
\end{array}=\begin{array}{c}
\Qcircuit @C=1em @R=1em {
&  \gate{R_y(\boldsymbol{\theta}_1)} & \gate{R_z(\boldsymbol{\theta}_6)} & \qw\\
& \gate{R_y(\boldsymbol{\theta}_2)} & \gate{R_z(\boldsymbol{\theta}_7)} & \qw\\
&\gate{R_y(\boldsymbol{\theta}_3)} & \gate{R_z(\boldsymbol{\theta}_8)} &\qw \\
&  \gate{R_y(\boldsymbol{\theta}_4)} &  \gate{R_z(\boldsymbol{\theta}_9)} &\qw\\
&  \gate{R_y(\boldsymbol{\theta}_5)} & \gate{R_z(\boldsymbol{\theta}_{10})} &\qw\\
}
\end{array}\]
\caption{Parameterized Pauli rotations acting on the qubits.}
\label{fig:RyRz}
\end{figure}

\begin{figure}[H]
\[\begin{array}{c}
\Qcircuit @C=1em @R=1em {
& \multigate{4}{ent} & \qw\\
&        \ghost{ent} & \qw\\
&        \ghost{ent} & \qw\\
&        \ghost{ent} & \qw\\
&        \ghost{ent} & \qw\\}
\end{array}
= 
\begin{array}{c}
\Qcircuit @C=1em @R=1em {
& \ctrl{1} & \qw      & \qw        & \qw      & \qw \\
& \targ    & \ctrl{1} & \qw        & \qw      & \qw \\
& \qw      & \targ    & \ctrl{1}   & \qw      & \qw \\
& \qw      & \qw      & \targ      & \ctrl{1} & \qw \\
& \qw      & \qw      & \qw        & \targ    & \qw \\
}
\end{array}\]
\caption{A ladder of linearly connected CNOT-gates that establishes entanglement in the ansatz.}
\label{fig:entanglement}
\end{figure}

The ansatz shown in \autoref{fig:fivequbitansatz} was inspired by a similar design proposed by \cite{expressivity}. They showed that this design quickly approximate random Haar unitaries, at the cost of relatively few CNOT operations. Since we want to approximate general quantum algorithms on noisy hardware, this design was a natural choice. The circuit depth of the ansatz is 26, but note that this depth can change when the ansatz is transpiled to respect a specific hardware. See \autoref{sec:CircuitDepths} for a definition of circuit depth. The parameters of the ansatz are initialized randomly as $\boldsymbol{\theta}_i \sim U[-\pi, \pi ]$, which coincides with the period of Pauli rotations. 

\subsection{Algorithm}
In algorithm \ref{alg:CompressionAlgorithm} we can see how RVQC is implemented.

\begin{algorithm}[H]
 \KwInput{ \\
 $\{U_k\}_{k=1,\cdots, N}$ - A circuit to be compiled, divided into $N$ parts (steps), represented by the unitary $U = U_N \cdots U_1$.\\ 
 $V(\boldsymbol{\theta})$ - A variational ansatz.\\
 $T$ - The number of training iterations. \\
 $\lambda$ - A learning rate for the gradient descent algorithm. }
 \KwOutput{\\
 $V (\boldsymbol{\theta}^{(N)})$ - A unitary approximation for the target circuit, $V (\boldsymbol{\theta}^{(N)})\ket{\boldsymbol{0}} \approx U_N \cdots U_1 \ket{\boldsymbol{0}}$. }
 \For{$k=1$ \KwTo $N$}
    {
    $\boldsymbol{\theta}^{(k)}_i \sim uniform(-\pi, \pi), \forall i$. \\
    calculate $C(\boldsymbol{\theta}^{(k)})$ in \autoref{eq:CostFunction} with the circuit in \autoref{fig:CostFunction}.\;
    \For{$t=1$ \KwTo $T$}{
        $\boldsymbol{\theta}^{(k)} \gets \boldsymbol{\theta}^{(k)} - \lambda \nabla_{\boldsymbol{\theta}^{(k)}} C(\boldsymbol{\theta}^{(k)})$  (or other gradient-based method, e.g., Adam)\; 
        calculate $C(\boldsymbol{\theta}^{(k)})$ in \autoref{eq:CostFunction} with the circuit in \autoref{fig:CostFunction}.\;
        }
    
    }
\caption{Recursive Variational Quantum Compiling (RVQC)}
\label{alg:CompressionAlgorithm}
\end{algorithm}

Since the variational ansatzes used in this work consists of only single-qubit Pauli rotations and non-parameterized multi-qubit gates, the gradients of the cost function with respect to the variational parameters are estimated using the parameter-shift rule \cite{PhysRevA.99.032331}. The parameters are then updated using the Adam gradient-based optimization method \cite{ADAM}.

\subsection{Noise Model}\label{sec:NoiseModel}
We utilized the FakeSantiago backend from Qiskit \cite{Qiskit}, which is a noise model mimicking the IBM Santiago $5$-qubit device. The basis gates for this device are CNOT, $I$, $R_z(\theta)$, $\sqrt{X}$ and $X$. Further, the qubits are linearly connected. In the model, the initial state is assumed ideal, and the measurements errors are assumed to be classical read-out errors. The qubits are subject to amplitude damping and depolarization. For more details, see \autoref{sec:Santiago Noise Model}

\subsection{Performance metrics}

We want an appropriate performance metrics to compare RVQC with VQC and with running the target circuit without compilation. We define a target pure density state, 
\begin{equation}
\label{eq:TargetState}
    \rho_I(U) = U\ket{\boldsymbol{0}}\bra{\boldsymbol{0}}U^\dagger,
\end{equation}
where $U$ is the circuit we wish to compile. A successful compilation should result in a state close to \autoref{eq:TargetState} under some measure. We also define a sequence of intermediate states,
\begin{equation}
\label{eq:IntermediateState}
    \rho_I(U^{(k)}) = U_k\cdots U_2 U_1\ket{\boldsymbol{0}}\bra{\boldsymbol{0}}U_1^\dagger U_2^\dagger \cdots U_k^\dagger,
\end{equation}
i.e., the state after evaluating the first $k$ sub-circuits of the full circuit $U$ on an ideal quantum computer. This state is of interest as we wish to evaluate our compilation during intermediate steps.

We want to compare the target states in \autoref{eq:TargetState} and \autoref{eq:IntermediateState} with our optimized ansatzes, with and without noise. We define, 

\begin{equation}
    \rho_I(V(\boldsymbol{\theta}^{(k)}); U^{(k)}) = V(\boldsymbol{\theta}^{(k)})\ket{\boldsymbol{0}}\bra{\boldsymbol{0}}V^\dagger(\boldsymbol{\theta}^{(k)}), 
\end{equation}
where the ansatz $V(\boldsymbol{\theta}^{(k)})$ has been optimized with VQC or RVQC to approximate the target circuit $U^{(k)}$. Equivalently, $\rho_N(U^{(k)})$ and $\rho_N(V(\boldsymbol{\theta}^{(k)}); U^{(k)})$ are the resulting mixed states from evaluation under non-unitary effects, like a simulated noise model.

A natural measure for similarity between quantum states is the Quantum State Fidelity \cite{fidelity}, with the usual definition,

\begin{equation}\label{eq:QSF}
    QSF[\rho_1, \rho_2] = |Tr(\sqrt{\sqrt{\rho_2}\rho_1\sqrt{\rho_2}})|^2.
\end{equation}

For any quantum states $\rho_1, \rho_2$, we have that $QSF(\rho_1, \rho_2) \in [0,1]$, with $QSF(\rho_1, \rho_2) = 1$ if and only if $\rho_1 =\rho_2$. To establish a comparative worse-case scenario, we compute the fidelity, 
\begin{equation}\label{eq:circuit_fid}
    F_N[U^{(k)}] = QSF[\rho_N(U^{(k)}), \rho_I(U^{(k)})],
\end{equation} i.e., the fidelity between the uncompiled first $k$ sub-circuits evaluated with noise, compared to the corresponding ideal result. As the circuit is very deep, we expect this to establish a "poor performance" comparison. We then compute the fidelities, 
\begin{equation}\label{eq:ansats_fid}
\begin{split}
    F_I[V(\boldsymbol{\theta}^{(k)}); U^{(k)}] = QSF[\rho_I(V(\boldsymbol{\theta}^{(k)}); U^{(k)}), \rho_I(U^{(k)})],\\ F_N[V(\boldsymbol{\theta}^{(k)}); U^{(k)}] = QSF[\rho_N(V(\boldsymbol{\theta}^{(k)}); U^{(k)}), \rho_I(U^{(k)})],
\end{split}    
\end{equation}
i.e., the ideal and noisy fidelities of the ansatzes approximating different number of sub-circuits $U^{(k)}$. Note that we can write $F_I[V(\boldsymbol{\theta}); U]$ and $F_N[V(\boldsymbol{\theta}); U]$ for VQC as there are no intermediate steps in this algorithm.

\begin{comment}
To compare the performance of the optimized ansatz, we compare the quantum state fidelity between the resulting state $\rho_1 = V(\boldsymbol{\theta})\ket{\boldsymbol{0}}\bra{\boldsymbol{0}}V(\boldsymbol{\theta})^\dagger$ and the target state $\rho_2 =  U\ket{\boldsymbol{0}}\bra{\boldsymbol{0}}U^\dagger$. The quantum state fidelity follows the usual definition
\begin{equation}
    QSF(\rho_1, \rho_2) = |Tr(\sqrt{\sqrt{\rho_2}\rho_1\sqrt{\rho_2}})|^2.
\end{equation}
For any quantum states $\rho_1, \rho_2$, we have that $QSF(\rho_1, \rho_2) \in [0,1]$, with $QSF(\rho_1, \rho_2) = 1$ if and only if $\rho_1 =\rho_2$. When $V(\boldsymbol{\theta})$ is evaluated as a unitary operator, the resulting state $\rho_1$ will be a pure state. The resulting fidelity will then be referred to as \textit{ideal} fidelity. If evaluated as a non-unitary operator, for example under the influence of a simulated noise model, $\rho_1$ will be in general a mixed state, and the resulting fidelity will be referred to as \textit{noisy} fidelity. In both cases, the target state $\rho_2$ will be evaluated ideally, without any noise. 

Finally, we establish a comparative worst-case scenario where the  full circuit $U$ is evaluated with a noise model, and the fid. 
\end{comment}

\section{Results and Discussion}
\label{sec:results}
We compare RVQC against standard VQC on a compiling benchmark consisting of several randomly generated five-qubit circuits, each with an approximate total depth of 1000. The exact circuit depths can be seen in \autoref{sec:CircuitDepths}. Deep random circuits often do not exhibit a specific structure, making the compilation a non-trivial task. We generated the random circuits by uniformly sampling gates from the IBM Santiago basis gate set (see \autoref{sec:NoiseModel}). The qubits on which the gates acted were selected randomly, ensuring that CNOT gates were only applied between connected qubits. Each full circuit $U$ was constructed by combining five independent random subcircuits, $\{U_k\}_{k=1,2,\dots, 5}$, each with a depth of 200. The full circuit is thus constructed as $U = U_5 U_4 \cdots U_1$, resulting in a total depth of approximately 1000. The subcircuits, $\{U_k\}_{k=1,2,\dots, 5}$, represents the divisions of the full circuit $U$ used in the RVQC algorithm.  We then utilized both VQC and RVQC to compile $U$, using \autoref{fig:fivequbitansatz} as our ansatz. Note that transpiling this ansatz to the IBM Santiago device results in a circuit with a total circuit depth of 36.

\begin{figure}[H]
    \centering
    \includegraphics[width=\linewidth]{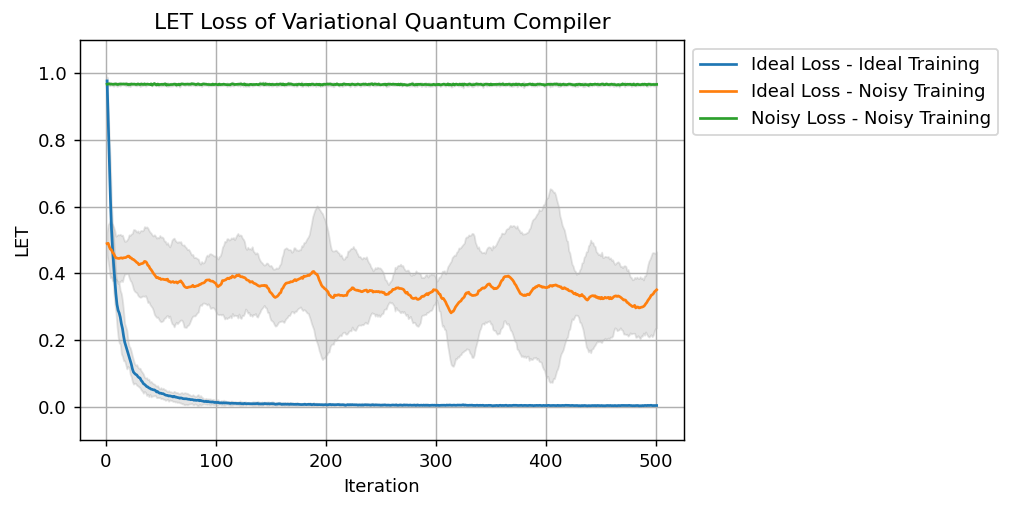}
    \caption{Training loss (LET \autoref{eq:CostFunction}) as a function of iteration for VQC on a randomly generated five-qubit circuit of approximately depth 1000. The fill-between marks two standard deviations. Ideal/noisy loss was evaluated on a simulation of ideal/noisy quantum hardware. Ideal/noisy training means that the VQC training was done on a simulation of ideal/noisy quantum hardware. The ideal and noisy fidelity between the resulting ansatz, $V(\boldsymbol{\theta})$, and the full circuit, $U$, was $F_I[V(\boldsymbol{\theta}); U] = 0.34 \pm 0.28$ and $F_N[V(\boldsymbol{\theta}); U] = 0.05 \pm 0.09$, respectively.}
    \label{fig:lossnoisyidealfivequbit}
\end{figure}

During compilation, VQC was allowed to optimize for 500 iterations, whereas RVQC used 100 iteration for each of the five subcircuits. This ensured a fair comparison, with a total of 500 iterations for each method. The Adam optimizer with a learning rate of $\lambda = 0.1$ was used in both cases, and standard deviations were computed across the five independent experiments. First, VQC was optimized by training it on Qiskits noiseless statevector simulator, and then by training it on Qiskits noise model of the IBM Santiago five-qubit machine \cite{Qiskit} with 8192 shots per circuit. The results can be seen in \autoref{fig:lossnoisyidealfivequbit}.

Figure \ref{fig:lossnoisyidealfivequbit} shows that when training on the ideal statevector simulator, VQC was able to converge to a satisfactory low loss value of $0.0038 \pm 0.0007$. However, it was not able to converge when trained with the noise model of the IBM Santiago device. Furthermore, the loss evaluated with the ideal and noisy simulation shows no obvious coinciding minima when performing the training on the noise model. Despite the OPR property, the noise is sufficiently large so that convergence to optimal, or even suitable, parameters is unfeasible. In practice, the gradient is too noisy to make meaningful updates beyond random steps in parameters space, even when using Adam in an effort to mitigate this. 

\begin{figure}[H]
    \centering
    \includegraphics[width=\linewidth]{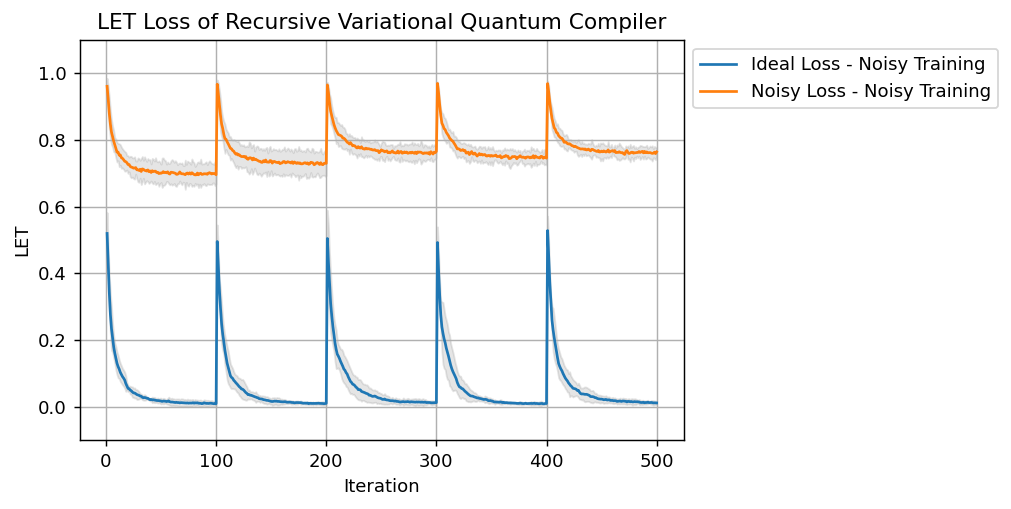}
    \caption{LET loss against iteration for the recursive variational quantum compiler. The loss for each individual sub-circuit is showed consecutively in the plot. The fill-between marks two standard deviations. Ideal/noisy loss means that the loss is evaluated on an ideal/noisy simulation of a quantum device. Noisy training means that the training is performed on an noisy simulation of a quantum device.}
    \label{fig:lossRVQC}
\end{figure}

We repeat the compilation of the random circuits using RVQC with the same noise model. \autoref{fig:lossRVQC} shows the progressive LET loss for each individual sub-circuit, showed consecutively in the plot.
We see that even though the noise prohibited the VQC to converge to a useful minimum, the RVQC is able to converge for each individual sub-circuit.

\begin{figure}[H]
    \centering
    \includegraphics[width=\linewidth]{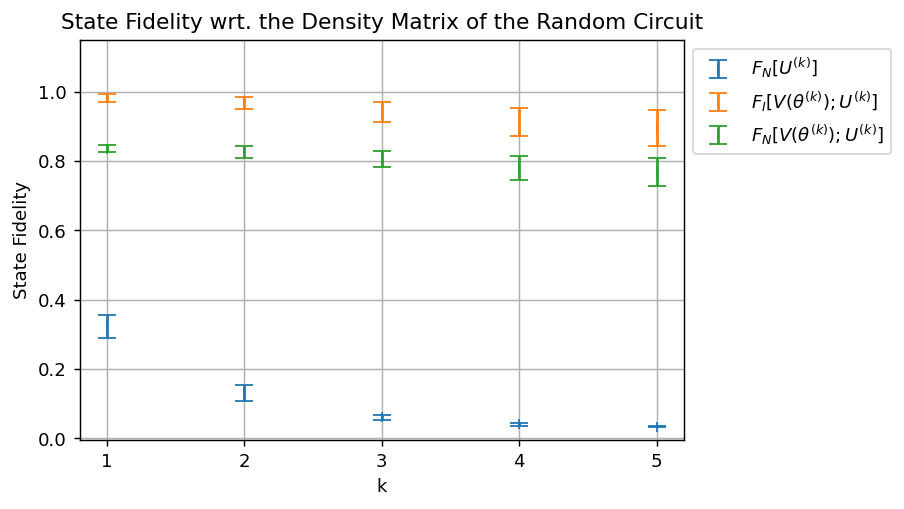}
    \caption{Quantum State Fidelity (\autoref{eq:ansats_fid}) as a function of the number of sub-circuits $k$, when optimizing with RVQC on sub-circuits $U^{(k)}$. Optimization is done under the influence of simulated noise. The target circuits are five independently sampled random quantum circuit with approximate depth of 1000, each split into five sub-circuits of depth 200.  Both the ideal (orange) and noisy (green) fidelity is computed with \autoref{eq:ansats_fid}. Two standard deviations are provided. RVQC is optimzied for 100 iterations on each sub-circuit, for a total of 500 iterations for the full circuit. As a worse-case comparison, the noisy fidelity \autoref{eq:circuit_fid} is computed for the uncompiled target circuit.}
    \label{fig:FiveQubitStateFidelity}
\end{figure}

The state fidelity between the RVQC at step $k$ and the first $k$ parts of the random circuits (see \autoref{eq:ansats_fid}) was calculated after each step. The results can be seen in \autoref{fig:FiveQubitStateFidelity}.

Figure \ref{fig:FiveQubitStateFidelity} shows that the noisy execution of the full random circuit yielded a fidelity of $F_N[U^{(5)}] = 0.034 \pm 0.003$, showing that some form of quantum compiling is necessary in order to achieve any meaningful results. Indeed, RVQC trained on the noisy device was able to compress the target circuit with an ideal and noisy fidelity of $F_I[V(\boldsymbol{\theta}^{(5)}); U^{(5)}] = 0.90 \pm 0.05$ and $F_N[V(\boldsymbol{\theta}^{(5)}); U^{(5)}] = 0.77 \pm 0.04$, respectively. In contrast, ordinary VQC was seen to not converge in \autoref{fig:lossnoisyidealfivequbit}, making RVQC the preferred choice for this example. By splitting the full random circuits into smaller parts, RVQC is able to decrease the amount of noise. This ensures in turn that the optimization procedure is accomplishable in 500 iterations.

A summarization of our results can be seen in \autoref{tab:FidelityTable}.

\begin{table}[H]
\centering
\caption{Ideal and noisy fidelity for RVQC and VQC. The mean across the five experiments is listed, along with two standard deviations. We include the fidelity between the full random circuit on the noise model and the actual density matrix for said circuit. }
\label{tab:FidelityTable}
\begin{tabular}{lcc}
\toprule
\textbf{Method} & \textbf{Noisy Fidelity} & \textbf{Ideal Fidelity} \\
\midrule
VQC & $0.05 \pm 0.09$ & $0.34 \pm 0.28$ \\
RVQC & $0.77 \pm 0.04$ & $0.90 \pm 0.05$ \\
Full Circuit & $0.034 \pm 0.003$ & 1 \\
\bottomrule
\end{tabular}
\end{table}

We argue that the convergence of the RVQC algorithm is due to the results presented in Section \ref{subsec:preventingExponentialGradientDecay}. The compilation by splitting approach yielded a gradient with a sufficient magnitude for convergence within the number of epochs, while the VQC algorithm did not.

\section{Conclusion}
\label{sec:Conclusion}
Five random five-qubit circuits of depth 1000 were compiled into parameterized ansatzes of depth 36 by utilizing RVQC and VQC. The circuits were split into $N=5$ parts when applying RVQC. It was shown that when optimizing for a total of 500 iterations on a simulation of a noisy quantum device, VQC was unable to converge while RVQC was able to train the ansatzes with a fidelity of $0.90 \pm 0.05$ with respect to the random circuits. The ansatzes were sufficiently shallow such that their density matrices on the noise model had an average fidelity of $0.77 \pm 0.04$ with respect to the actual density matrices of the random circuits. For comparison, the density matrices for the full random circuits on the noise model had a corresponding fidelity of only $0.034 \pm 0.003$.
We argue that the success of the RVQC algorithm is due to mitigation of noise-induced barren plateaus.

\subsection{Further Work}
The RVQC algorithm depends on the ansatz being expressive enough to successfully complete all of the $N$ steps. Meanwhile, it is expected that the required expressiveness will be varying with each step. An adaptive ansatz approach \cite{bilkis2022semiagnostic,grimsleyadaptiveansatz} thus seems like a natural extension of this work.

Research has shown that for a wide class of parameterized quantum circuits, the probability of non-zero gradients vanishes exponentially in the number of qubits \cite{barrenplateusinquantumneuralnetworks}, meaning that one has to choose ansatzes with care. Several ways to mitigate this problem have been proposed. For example, a cost function dependent on local observables will lead to only polynomially vanishing gradients if one uses an alternating layered ansatz composed of blocks forming local 2-designs, and the circuit depth does not exceed $\mathcal{O}(\log n)$ \cite{CostfunctionDependentBarrenPlateus}. It would be interesting to find out if such ansatzes can be flexible enough to compress deep quantum circuits.

The RVQC algorithm aims to perform FISC, as opposed to FUMC. Note, however, that RVQC can easily be extended to the FUMC case by changing to the appropriate cost function. 

Finally, although the RVQC algorithm was tested out on the IBM Santiago noise model, it has not yet been tested on any real quantum devices. It would be interesting to see if the method is applicable with realistic noise.

\section{Acknowledgements}
We would like to thank Morten Hjorth-Jensen for sharing his knowledge and for engaging in productive discussions. His insights have been helpful and contributed to the progress of our work.

\bibliographystyle{unsrtnat}
\bibliography{template}  

\appendix
\section{Santiago Noise Model}\label{sec:Santiago Noise Model}
In this section, we present the noise characteristics of the Santiago quantum processor, focusing on the various sources of errors.

\subsection*{Readout Errors}
Table \ref{tab:ReadoutError} summarizes the readout error probabilities for each qubit. These probabilities indicate the likelihood that the measurement result deviates from the expected value due to noise during the readout process.

\begin{table}[H]
    \centering
    \caption{Readout error probabilities for each qubit.}
    \label{tab:ReadoutError}
    \begin{tabular}{cccc}
    \toprule
    \textbf{Qubit} & \textbf{P(j|0)} & \textbf{P(j|1)} \\
    \midrule
    0 & $[0.9936, 0.0064]$ & $[0.0202, 0.9798]$ \\
    1 & $[0.9892, 0.0108]$ & $[0.0180, 0.9820]$ \\
    2 & $[0.9890, 0.0110]$ & $[0.0250, 0.9750]$ \\
    3 & $[0.9848, 0.0152]$ & $[0.0160, 0.9840]$ \\
    4 & $[0.9958, 0.0042]$ & $[0.0150, 0.9850]$ \\
    \bottomrule
    \end{tabular}
\end{table}

\subsection*{T1 and T2 Times}
Table \ref{tab:T1T2} shows the $T_1$ and $T_2$ times for each qubit, which represent the relaxation and dephasing times, respectively.
\begin{table}[H]
    \centering
    \caption{T1 and T2 times in seconds for each qubit.}
    \label{tab:T1T2}
    \begin{tabular}{ccc}
    \toprule
    \textbf{Qubit} & \textbf{T1} & \textbf{T2} \\
    \midrule
    0 & $7.474 \times 10^{-5}$ & $1.286 \times 10^{-4}$ \\
    1 & $1.541 \times 10^{-4}$ & $9.246 \times 10^{-5}$ \\
    2 & $1.156 \times 10^{-4}$ & $9.927 \times 10^{-5}$ \\
    3 & $1.477 \times 10^{-4}$ & $9.332 \times 10^{-5}$ \\
    4 & $1.280 \times 10^{-4}$ & $1.219 \times 10^{-4}$ \\
    \bottomrule
    \end{tabular}
\end{table}

\subsection*{Gate Error Rates}
Table \ref{tab:GateError} provides the error rates for different gate operations. The error rate refers to the probability of an error occurring when applying the respective gate to one or more qubits.
\begin{table}[H]
    \centering
    \caption{Gate error rates for each gate and qubit pair.}
    \label{tab:GateError}
    \begin{tabular}{cccc}
    \toprule
    \textbf{Gate} & \textbf{Qubits} & \textbf{Error Rate} \\
    \midrule
    ID  & [0] & $2.067 \times 10^{-4}$ \\
    ID  & [1] & $1.658 \times 10^{-4}$ \\
    ID  & [2] & $2.582 \times 10^{-4}$ \\
    ID  & [3] & $1.838 \times 10^{-4}$ \\
    ID  & [4] & $1.781 \times 10^{-4}$ \\
    RZ  & [0] & $0$ \\
    RZ  & [1] & $0$ \\
    RZ  & [2] & $0$ \\
    RZ  & [3] & $0$ \\
    RZ  & [4] & $0$ \\
    SX  & [0] & $2.067 \times 10^{-4}$ \\
    SX  & [1] & $1.658 \times 10^{-4}$ \\
    SX  & [2] & $2.582 \times 10^{-4}$ \\
    SX  & [3] & $1.838 \times 10^{-4}$ \\
    SX  & [4] & $1.781 \times 10^{-4}$ \\
    X   & [0] & $2.067 \times 10^{-4}$ \\
    X   & [1] & $1.658 \times 10^{-4}$ \\
    X   & [2] & $2.582 \times 10^{-4}$ \\
    X   & [3] & $1.838 \times 10^{-4}$ \\
    X   & [4] & $1.781 \times 10^{-4}$ \\
    CX  & [4, 3] & $5.200 \times 10^{-3}$ \\
    CX  & [3, 4] & $5.200 \times 10^{-3}$ \\
    CX  & [2, 3] & $5.720 \times 10^{-3}$ \\
    CX  & [3, 2] & $5.720 \times 10^{-3}$ \\
    CX  & [2, 1] & $6.886 \times 10^{-3}$ \\
    CX  & [1, 2] & $6.886 \times 10^{-3}$ \\
    CX  & [0, 1] & $6.300 \times 10^{-3}$ \\
    CX  & [1, 0] & $6.300 \times 10^{-3}$ \\
    \bottomrule
    \end{tabular}
\end{table}

\subsection*{Gate Durations}
Table \ref{tab:GateDurations} lists the durations of various gate operations, including both single-qubit and two-qubit gates. The duration of a gate impacts the overall execution time of quantum circuits, with two-qubit gates typically requiring longer durations.
\begin{table}[H]
    \centering
    \caption{Gate durations in seconds for each gate and qubit pair.}
    \label{tab:GateDurations}
    \begin{tabular}{cccc}
    \toprule
    \textbf{Gate} & \textbf{Qubits} & \textbf{Duration} \\
    \midrule
    ID  & [0] & $3.556 \times 10^{-8}$ \\
    ID  & [1] & $3.556 \times 10^{-8}$ \\
    ID  & [2] & $3.556 \times 10^{-8}$ \\
    ID  & [3] & $3.556 \times 10^{-8}$ \\
    ID  & [4] & $3.556 \times 10^{-8}$ \\
    RZ  & [0] & $0$ \\
    RZ  & [1] & $0$ \\
    RZ  & [2] & $0$ \\
    RZ  & [3] & $0$ \\
    RZ  & [4] & $0$ \\
    SX  & [0] & $3.556 \times 10^{-8}$ \\
    SX  & [1] & $3.556 \times 10^{-8}$ \\
    SX  & [2] & $3.556 \times 10^{-8}$ \\
    SX  & [3] & $3.556 \times 10^{-8}$ \\
    SX  & [4] & $3.556 \times 10^{-8}$ \\
    X   & [0] & $3.556 \times 10^{-8}$ \\
    X   & [1] & $3.556 \times 10^{-8}$ \\
    X   & [2] & $3.556 \times 10^{-8}$ \\
    X   & [3] & $3.556 \times 10^{-8}$ \\
    X   & [4] & $3.556 \times 10^{-8}$ \\
    CX  & [4, 3] & $3.413 \times 10^{-7}$ \\
    CX  & [3, 4] & $3.769 \times 10^{-7}$ \\
    CX  & [2, 3] & $3.769 \times 10^{-7}$ \\
    CX  & [3, 2] & $4.124 \times 10^{-7}$ \\
    CX  & [2, 1] & $5.689 \times 10^{-7}$ \\
    CX  & [1, 2] & $6.044 \times 10^{-7}$ \\
    CX  & [0, 1] & $5.262 \times 10^{-7}$ \\
    CX  & [1, 0] & $5.618 \times 10^{-7}$ \\
    \bottomrule
    \end{tabular}
\end{table}

\section{Circuit Depths}\label{sec:CircuitDepths}

Circuit depth is the number of sequential layers of quantum gates applied to qubits in a quantum circuit. It represents the longest sequence of dependent operations, or time steps, required to complete a computation \cite{NielsenChuang2010}. Circuit depth is critical since quantum computers are sensitive to errors as time increases, making shallow circuits more reliable.

In order to calculate the circuit depth, we need to identify parallel and sequential gates: Gates acting on different qubits simultaneously (in parallel) belong to the same layer and do not increase the depth. Gates applied sequentially to the same qubit or dependent on previous operations create new layers and add to the depth. The circuit depth is then determined by the longest sequence of dependent operations.

Table \ref{tab:circuit_depths} displays the depths of five different circuits used for Variational Quantum Compiling (VQC) and Recursive Variational Quantum Compiling (RVQC). The depths slightly differ due to the random nature of the circuit generation.

\begin{table}[H]
\centering
\caption{Circuit depths of the five random circuits utilized for VQC and RVQC. The slightly different circuit depths is a result of adding together five random 200 depth circuits. }
\label{tab:circuit_depths}
\begin{tabular}{|c|c|}
\hline
\textbf{Circuit} & \textbf{Depth} \\ \hline
Circuit 1 & 1000 \\ \hline
Circuit 2 & 999  \\ \hline
Circuit 3 & 997  \\ \hline
Circuit 4 & 999  \\ \hline
Circuit 5 & 997  \\ \hline
\end{tabular}
\end{table}

\end{document}